\documentclass[aps,prd,twocolumn,nofootinbib,superscriptaddress]{revtex4-1}
\usepackage[utf8]{inputenc}
\usepackage{graphicx}
\usepackage{subfigure}
\usepackage{datatool}
\usepackage{amsmath,amssymb}
\usepackage[scientific-notation=true]{siunitx}
\usepackage{lineno}
\usepackage{natbib}
\usepackage[dvipsnames]{xcolor}
\usepackage{xspace}

\DTLsetseparator{,}

\def\gw{gravitational wave\xspace}
\def\ifos{interferometers\xspace}
\def\gwh{gravitational-wave\xspace}
\def\gws{gravitational waves\xspace}
\def\dm{dark matter\xspace}
\def\dmh{dark-matter\xspace}

\def\smh{standard-model\xspace}
\def\uldm{ultralight dark matter\xspace}
\def\uldmh{ultralight dark-matter\xspace}
\def\pbhs{primordial black holes\xspace}

\newcommand{\TFFT}{T_\text{FFT}}
\newcommand{\tfftmax}{T_\text{FFT,max}}

\newcommand{\bea}{\begin{eqnarray}}
\newcommand{\eea}{\end{eqnarray}}
\newcommand{\be}{\begin{equation}}
\newcommand{\ee}{\end{equation}}

\newcommand{\rhoDM}{\rho_{\text{DM}}}

\newtoggle{fullauthorlist}
\toggletrue{fullauthorlist}
\newtoggle{endauthorlist}
\toggletrue{endauthorlist}

\begin{document}

\title{Distinguishing between dark-matter interactions  with gravitational-wave detectors }

\author{Andrew L. Miller}
\email{andrew.miller@uclouvain.be}
\affiliation{Université catholique de Louvain, B-1348 Louvain-la-Neuve, Belgium}

\author{Francesca Badaracco}
\email{francesca.badaracco@uclouvain.be}
\affiliation{Université catholique de Louvain, B-1348 Louvain-la-Neuve, Belgium}
\author{Cristiano Palomba}
\affiliation{INFN, Sezione di Roma, I-00185 Roma, Italy}
\email{cristiano.palomba@roma1.infn.it}


\begin{abstract}
Ground-based \gwh \ifos could directly probe the existence of \uldm ($\mathcal{O}(10^{-14}-10^{-11})$ eV/$c^2$) that couples to \smh particles in the detectors.
Recently, many techniques have been developed to extract a variety of potential \dmh signals from noisy \gwh data; however, little effort has gone into ways to distinguish between types of \dm that could directly interact with the \ifos.
In this work, we employ the Wiener filter to follow-up candidate \dmh interaction signals. 
The filter captures the stochastic nature of these signals, and, in simulations, successfully identifies which type of \dm interacts with the \ifos. The power of this method to distinguish between different types of \dm comes from different coupling mechanisms that result in different power spectra, as well as different correlations between detectors spread across the earth. We apply the Wiener filter to outliers that remained in the LIGO/Virgo/KAGRA search for dark photons in data from the most recent observing (O3) \cite{LIGOScientific:2021odm}, and show that they are consistent with noise disturbances. Our proof-of-concept analysis demonstrates that the Wiener filter can be a powerful technique to confirm or deny the presence of \dmh interaction signals in \gwh data, and distinguish between scalar and vector \dmh interactions.
\end{abstract}

\maketitle
\section{Introduction}\label{sec:intro}

The existence of \dm has puzzled scientists for the last few decades. While ample evidence supports an invisible type of matter that moves the stars in our galaxy around faster than we would expect based on visible matter \cite{Corbelli:1999af}, that gravitationally lenses light in the Bullet cluster \cite{Clowe:2003tk}, and that explains anisotropies in the cosmic microwave background power spectrum \cite{Sugiyama:1994ed}, the underlying nature of \dm has eluded our understanding. Theories of beyond \smh physics allow \dm to have a mass as light as $\sim 10^{-22}$ eV$/c^2$ or as heavy as $\sim1$ PeV$/c^2$ \cite{Feng:2010gw}. Additionally, \dm could be macroscopic and composed in part or completely of \pbhs \cite{Green:2020jor}. 
To search for \dm in such a wide parameter space, different experiments have been designed, some that probe \dm via its direct interaction (scattering) with \smh particles \cite{XENON:2016jmt,SuperCDMS:2018mne,DarkSide:2018ppu}, and others that look for \dm indirectly via electromagnetic signatures resulting from the annihilation or decay of \dmh particles \cite{Gaskins:2016cha}. 

Though not constructed for the specific reason to detect \dm, \gwh interferometers, such as LIGO \cite{2015CQGra..32g4001L}, Virgo \citep{2015CQGra..32b4001A}, and KAGRA \cite{Aso:2013eba}, offer an innovative and competitive way to search for \uldm in the mass range $\mathcal{O}(10^{-14}-10^{-11})$ eV/$c^2$. These detectors rely on high-precision measurements of the positions of the mirrors in each arm of the laser interferometer that would follow a path in spacetime carved out by a passing \gw \cite{maggiore2008gravitational}. In our work, though, we do not look for a signature of \gws, but for one of a direct interaction of \dmh particles with components of the \gwh interferometers. Thus, LIGO, Virgo, and KAGRA become similar to particle physics, direct-detection experiments.

Recently, searches for different types of scalar and vector \uldm have been performed. The analysis of GEO600 data \cite{Dooley:2015fpa} using a Logarithmic frequency axis Power Spectral Density (LPSD) method \cite{trobs2006improved,trobs2009improved} yielded competitive constraints on scalar, dilaton dark matter \cite{Stadnik2015a,Stadnik2015b,Stadnik2016,grote2019novel} that could have coupled to electrons and photons in the beam splitter \cite{Vermeulen:2021epa}. Furthermore, constraints on vector dark matter, i.e. dark photons, were placed using data from the first \cite{guo2019searching} and third \cite{LIGOScientific:2021odm} observing runs of advanced LIGO/Virgo that surpassed upper limits from the Eöt-Wash \cite{Schlamminger:2007ht} and MICROSCOPE \cite{Berge:2017ovy} experiments by a few orders of magnitude at frequencies between $\sim100-1000$ Hz ($4\times 10^{-13}-4\times 10^{-12}$ eV/$c^2$). 
The existence of \uldm has also been constrained by searching for \gws from depleting boson clouds around black holes \cite{palomba2019direct,Sun:2019mqb,isi2019directed,d2018semicoherent,Ng:2020ruv,Zhu:2020tht,Tsukada:2018mbp}, and by analyzing mergers, e.g. GW190521 \cite{LIGOScientific:2020iuh}, which was shown to be consistent with the merger of complex vector boson stars \cite{CalderonBustillo:2020srq}.

Though the field of direct \dmh detection with \gwh interferometers is blossoming, a key question remains unanswered: in the event of a detection, will it be possible to \emph{distinguish} among different \uldmh interaction models? As explained in the next section, \dm could be composed of scalar or vector particles that would both leave similar imprints on \gwh interferometers. Current analysis methods based on cross-correlation \cite{PierceRilesZhao2018} and excess power \cite{Miller:2020vsl} allow for the detection of \emph{a} \dmh particle, but cannot determine \emph{which} \dmh particle has actually been observed, because these methods match their analysis coherence times to the coherence time of the \uldmh signal. This finite coherence time arises because individual particles in the \dmh wave packet travel with slightly different velocities that follow a Maxwell-Boltzmann distribution centered about the virial velocity $v_0$ of \dm: $v_0=220$ km/s\footnote{This quantity is the velocity at which \dm orbits the center of our Galaxy.}. In this work, we propose a new method based on the Wiener filter \cite{wiener1964extrapolation} to distinguish among the different types of interactions using their similar but distinct power spectra.

\section{Dark matter interaction models}\label{sec:dmints}

Different types of \dm would leave different signatures in \gwh interferometers. Regardless of the type of \dm, certain properties collectively characterize the nature of the signal \cite{Carney:2019cio}. First, the occupation number of \uldm is gigantic, $\mathcal{O}(10^{50})$, which implies that the wavefunctions of individual \dmh particles overlap. Second,  \uldm is cold, so the velocities of these \dmh particles follow a Maxwell-Boltzmann distribution, centered about the virial velocity. Third, \dm behaves as a classical sinusoidal field within its coherence time, oscillating at a frequency that is proportional to the \dmh mass. Fourth, when observing for longer than a coherence time, \uldm will not oscillate at a fixed frequency, but \emph{about} that frequency, with stochastic frequency variations $\Delta f/f\sim \mathcal{O}(v_0^2/c^2)\sim 10^{-6}$ \cite{PierceRilesZhao2018}.

In the following subsections, we describe the differences between ultralight scalar and vector \dm, and the specific ways in which these particles would generate a signal in \gwh detectors.

\subsection{Scalar dark matter}\label{subsec:scalar}

 Models for scalar, spin-0 \dm have received a lot of attention over the last few decades. Examples of these particles include the quantum chromodynamics (QCD) axion ~\cite{Peccei1977,Peccei1977PhRvD,Weinberg1978,Wilczek1978}, which appears naturally as a pseudo Nambu-Goldstone boson of a spontaneous global $U(1)$ symmetry breaking that also solves the strong charge-parity (CP) problem; and the dilaton, which can occur in multi-dimensional theories \cite{Preskill:1982cy,Abbott:1982af,Dine:1982ah,Cho:1998aa,Cho:2007cy,Arvanitaki:2014faa}. 
 
Axions could alter the phase velocities of circularly polarized photons in the laser beams traveling down each arm of the detector \cite{nagano2019axion}. Such a signal would actually be visible in other channels aside from the canonical \gwh one \cite{nagano2021axion}, and would require additional but simplistic optical components to measure the optical path difference
between $p$- and $s$- polarized light. In practice, linearly polarized light ($p$-) is inputted, and the axion causes polarization modulations, producing $s$- polarized light. Linearly polarized light can be expressed as a superposition of circularly polarized light.
 
On the other hand, dilaton-like \dm \cite{Stadnik2015a,Stadnik2015b,Stadnik2016} could change the mass of the electron and other physical constants, causing oscillations in the Bohr radius of atoms in various components of the interferometer \cite{grote2019novel}. In particular, the size and index of refraction of the beam splitter would oscillate over time; thus, light rays returning to the beam splitter from the interferometer cavities would traverse slightly different distances on the surface of the beam splitter, leading to a differential strain.

Such a scalar \uldmh field $\phi$ can be written as \cite{Arvanitaki:2014faa,Derevianko:2016vpm,Vermeulen:2021epa}:
\begin{equation}\label{eq:osc_field}
        \phi(t,\vec{r}) = \left(\frac{\hbar \sqrt{2 \rho_{\rm{DM}}}}{m_\phi c}\right) \cos\left(\omega_\phi t - \vec{k}_\phi \cdot \vec{r}\right),
\end{equation}
where $t$ is time, $\vec{r}$ is a position vector, $\hbar$ is Planck's reduced constant, $c$ is the speed of light, $\omega_\phi = 
(m_\phi c^2) / \hbar$ is the angular Compton frequency, $\vec{k}_\phi = 
(m_\phi\vec{v}_{\text{obs}} )/ \hbar $ is the wave vector, $m_\phi$ is the mass of the field, and $\vec{v}_{\text{obs}}$ is the velocity of the \dm relative to the observer. 

The Lagrangian $\mathcal{L}_\mathrm{int}$ for this scalar field is \cite{Vermeulen:2021epa}:

\begin{equation}\label{L_int}
    \mathcal{L}_\mathrm{int} \supset \frac{\phi}{\Lambda_\gamma} \frac{F_{\mu\nu}F^{\mu\nu}}{4}  - \frac{\phi}{\Lambda_e} m_e \bar{\psi}_e \psi_e,
\end{equation}
where $F_{\mu \nu} = \partial_\mu A_\nu - \partial_\nu A_\mu$ is the electromagnetic field tensor, $\psi_e$ and $\bar{\psi}_e$ are the standard-model electron field and its Dirac conjugate, and $\Lambda_\gamma$ and $\Lambda_e$ denote the scalar \dmh coupling parameters to the photon and electron, respectively. 

Such couplings would cause changes in the index of refraction and sizes of materials, and would lead to a differential displacement $\delta (L_x - L_y)$ on \gwh interferometers given by \cite{grote2019novel}:

\begin{equation}\label{eq:full_signal}
    \delta (L_x - L_y) \approx \left(\frac{1}{\Lambda_\gamma} + \frac{1}{\Lambda_e} \right)\left(\frac{n\,l\,\hbar\,\sqrt{2\,\rho_{\mathrm{DM}}}}{m_\phi\, c}\right)\cos\left(\omega_{\text{obs}} t\right),
\end{equation}
where $n$ and $l$ are the index of refraction and length of the beam splitter, respectively.


\subsection{Vector dark matter}\label{subsec:vector}

Dark matter could be composed of spin-1 particles, which we denote as the dark photon. The relic abundance of dark matter can be explained entirely by dark photons, which could arise from the misalignment mechanism \cite{nelson2011dark,arias2012wispy,graham2016vector}, parametric resonance or the tachyonic instability of a scalar field \cite{agrawal2020relic,pierce2019dark,bastero2019vector,dror2019parametric}, or from cosmic string network decays \cite{long2019dark}. Dark photons could couple directly to baryon or baryon-lepton number in the four primary
interferometer mirrors that serve as \gwh test masses, and exert a ``dark'' force on the mirrors, causing quasi-sinusoidal oscillations \cite{Miller:2020vsl,PierceRilesZhao2018}. 

 We formulate dark photons in an analogous way to ordinary photons: as having a vector potential with an associated dark electric field that causes a quasi-sinusoidal force on the mirrors in the interferometers. 

The vector potential for a single dark photon particle can be written as:

\be
\vec{A}=\left(\frac{\hbar \sqrt{2\rhoDM}}{m_A c^2}\frac{1}{\sqrt{\epsilon_0}}\right)\sin\left(\omega_A t -\vec{k}_A \cdot\vec{r}+ \Upsilon\right),
\label{numI}
\ee
where $\omega_A = 
(m_A c^2) / \hbar$ is the angular Compton frequency, $\vec{k}_A = 
(m_A\vec{v}_{\text{obs}} )/ \hbar $ is the wave vector, $m_A$ is the mass of the vector field, $\epsilon_0$ is the permittivity of free space and $\Upsilon$ is a random phase.

The Lagrangian $\mathcal{L}$ that characterizes the dark photon coupling to a number current density $J^\mu$ of baryons or baryons minus leptons is:

\begin{equation}
    \mathcal{L} = -\frac{1}{4\mu_0} F^{\mu \nu} F_{\mu \nu} + \frac{1}{2\mu_0} \left(\frac{m_A c }{\hbar}\right)^2 A^\mu A_\mu - \epsilon e J^\mu A_\mu, \label{lagrangian} \\
\end{equation}
where $F_{\mu\nu}$ now indicates the \emph{dark} electromagnetic field tensor, $\mu_0$ is the magnetic permeability in vacuum, $m_A$ is the dark photon mass, $A_\mu$ is the four-vector potential of the dark photon, $e$ is the electric charge, and $\epsilon$ is the strength of the particle/dark photon coupling normalized by the electromagnetic coupling constant.

Dark photons cause small motions of an interferometer's mirrors, and lead to an observable effect in two ways. Firstly, the mirrors are well-separated from each other and hence experience slightly different dark photon dark matter phases. Such a phase difference leads to a differential change of the arm length, suppressed by $v_0/c$. This effect is in fact a residual one: if the mirrors of current \gwh \ifos had different material
compositions from each other, then the signal induced from dark photons coupling to baryon-lepton number would be enhanced \cite{Michimura:2020vxn}. A simple relation between dark photon parameters and the effective strain $h_D$ can be written as \cite{PierceRilesZhao2018}:

\bea
\sqrt{\left\langle h_{D}^{2}\right\rangle} &=&C\frac{q}{M}\frac{v_{0}}{2\pi c^{2}} \sqrt{\frac{2\rho_{\mathrm{DM}}}{\epsilon_{0}}}\frac{e\epsilon}{f_{0}}, 
\label{h000}
\eea
where $q/M$ is the charge-to-mass ratio of the mirrors, $f_0$ is the frequency of the \dmh particle, and $C=\sqrt{2}/3$ is a geometrical factor obtained by averaging over all possible dark photon propagation and polarization directions.

Secondly, the common motion of the interferometer mirrors, induced by the dark photon dark matter background, can lead to an observable signal because of the finite travel time of the laser light in the interferometer arms. The light will hit the mirrors at different times during their common motions, and although the common motions do not change the instantaneous arm length, they can lead to a longer round-trip travel time for the light, equivalent to arm lengthening, and therefore an apparent differential strain~\cite{Morisaki:2020gui}. Similarly to equation \ref{h000}, the common motion induces an observable signal with an effective strain $h_C$ as:
\bea
\sqrt{\langle h_C^2\rangle}&=&\frac{\sqrt 3}{ 2} \sqrt{\langle h_D^2\rangle}\frac{2\pi f_0 L}{v_0}.
\label{h000p}
\eea
The interference between the two contributions to the strain averages to zero over time, which means that the total effective strain can be written as $\langle h_{{\rm total}}^2\rangle = \langle h_D^2\rangle+\langle h_C^2\rangle$.

\section{Wiener Filtering}\label{sec:method}.

The Wiener filter has been applied in the context of \gwh data analysis before, specifically in terms of reducing Newtonian noise \cite{Badaracco:2020qmm}, detecting a stochastic \gwh background \cite{Meyers:2020qrb}, and handling the presence of correlated noise \cite{Thrane:2014yza,Coughlin:2016vor,Coughlin:2018tjc}. Here, however, we apply the Wiener filter to a new problem: searching for \uldm.

Coherence times for \uldm in the mass range considered here range from hours to days, which are much less than the approximately one-year observation time of a ground-based \gwh detector. Thus, standard \uldmh searches \cite{Miller:2020vsl,PierceRilesZhao2018,guo2019searching,LIGOScientific:2021odm,Vermeulen:2021epa} must restrict their analysis coherence times (that is, the fast Fourier Transform length $\TFFT$) to match that of the \uldmh coherence time, and sum the power in these individual chunks. While these methods ensure that power remains within one frequency bin in each chunk, they mask the stochastic nature of the signal; thus, all \uldmh interactions with \gwh detectors would appear the same if detected by these methods. Thus, we propose to use Wiener filtering to extract different types of \uldmh signals from LIGO/Virgo/KAGRA data.

\subsection{Method Description}
 Typically, the Wiener Filter can be used to estimate signal parameters in the presence of noise, and its formulation is the basis of least square error applications (such as linear prediction and adaptive filters). The Wiener filter coefficients are calculated by minimizing the mean squared error between a finite impulse response (FIR) filter (Equation \ref{eq:WF-time-convolution}) and the target signal. The least square error filter theory assumes that the signals are stationary processes \citep{Vas2001}.
We can linearly reconstruct a signal from a multiple inputs (Multiple Input Single Output--MISO-- filter):
\begin{equation}
\begin{split}\label{eq:WF-time-convolution}
\hat{x}[m] &= \sum_{k = 0}^{P-1}w_{k_0}y_0[m-k] + ... + \sum_{k = 0}^{P-1}w_{k_N}y_N[m-k] =\\
&= \sum_{k = 0}^{P-1}\mathbf{w}^{T}_k\mathbf{y}[m-k].
\end{split}
\end{equation}
Here, $\hat{x}[m]$ is the estimated discrete target signal at time $m$ (the \uldmh signal in our case) and $y_i[m-k]$ are the $N$ discrete input signals collected from $N$ witness sensors at the time $m-k$ (the \gwh detectors). This is MISO linear filter of order $P$, which means that we use $P$ coefficients and $P$ past values of each witness signal, $y_i[m]$; $\mathbf{w}_k$ represents the k$^{th}$ vector of the Wiener filter coefficients, while $\mathbf{y}[m]$ is the vector containing the $N$ signals from all the witness channels. 
For the purposes of detecting and validating an \uldmh signal, we do not need to reconstruct it in the time domain; therefore, we can employ the Wiener filter in the {frequency} domain. In this way, the information regarding the order of the filter disappears. Indeed, Equation \ref{eq:WF-time-convolution} is a discrete convolution of functions with finite support, so we can apply the convolution theorem, which connects the $z$-Transforms of the filter input and output by means of the filter transfer function. For a filter described by Equation \ref{eq:WF-time-convolution}, it is always possible to work on the unit circle of the complex plane and express the relation among input and output through the discrete-time Fourier Transform \cite{smith2007introduction}. This means that we can apply the Fourier transform to a block of $M$ samples and rewrite Equation \ref{eq:WF-time-convolution} as:
\begin{equation}\label{eq:WF-freq-product}
\hat{X}(\omega) = \mathbf{W}^{T}(\omega)\mathbf{Y}(\omega) = \mathbf{Y}^{T}(\omega)\mathbf{W}(\omega),
\end{equation}
where both $\mathbf{W}(\omega)$ and $\mathbf{Y}(\omega)$ are $N$-dimensional vectors containing the Discrete Fourier transforms of the Wiener filter coefficients and the $N$ witness signals, respectively. The Wiener filter is then defined by the coefficients that minimize the ensemble average of the square error function, $E[e^*[m]e[m]]$, with $e[m] = x[m]-\hat{x}[m]$:
\begin{equation}\label{eq:square-error}
E\left[e^*e\right] = E\left[\left(X - \mathbf{Y}^{T}\mathbf{W}\right)^*\left(X - \mathbf{Y}^{T}\mathbf{W}\right)\right].
\end{equation} 
Minimizing $E[e^*[m]e[m]]$ with respect to the Wiener filter coefficients, we obtain its optimal value: 
\begin{equation}\label{eq:WF-coeff}
\mathbf{W} = \left(\bar{\mathbf{P}}_{YY}\right)^{-1}\mathbf{P}_{XY},
\end{equation}
where $\bar{\mathbf{P}}_{YY}$ is the $N$x$N$ matrix containing the cross-Power Spectral Densities (PSD) of the $N$ witness sensors while $\mathbf{P}_{XY}$ is the vector containing the cross-PSD between the $N$ witness sensors and the target signal model. With this result, we can rewrite $E[e^*e]$ to define the residual as the ensemble average of the least square error normalized by the target signal PSD ($P_{XX}(\omega) = E[X^*(\omega)X(\omega)]$): 
\begin{equation}\label{eq:WF-residual}
R(\omega) = 1 - \frac{\mathbf{P}^{\dag}_{XY}\bar{\mathbf{P}}_{YY}^{-1}\mathbf{P}_{XY}}{P_{XX}}.
\end{equation}
Given the properties of the PSD, $R(\omega)$ will be real valued and constrained between 0 and 1. In the context of this work, $X(\omega)$ represents the template of the \uldmh signal we will use to determine if the data $Y(\omega)$ contain the \uldmh signal. 

\subsection{Combining multiple detectors}\label{subsec:combin}
Assuming each detector's signal is the sum between the \uldmh signal and the detector's self noise, we can retrieve the self-noise limiting curve of the Wiener filter. We consider the ideal case in which correlations between the three detectors' noises, as well as between the noises and the signal, are negligible; thus, we can write the residual as: 
\begin{equation}
    R = \frac{1}{1 + \frac{1}{SNR_{L}^2 + SNR_{H}^2 + SNR_{V}^2}},
\end{equation}
where $SNR_i^2 = E[x,x]/E[n_i,n_i]$ is the signal-to-noise ratio (SNR) of each detector $i$. Thus using more detectors helps to enhance the total SNR and improves the residual. Having a detector with a much smaller SNR does not decrease the capability of the Wiener filter; however, it does not contribute much to lowering the residual relative to the other two detectors. Of course, any correlation between the signal and the detector's noise and between different detectors, can spoil the residual.      
\section{Results}\label{sec:result}
The Wiener filter requires a model for the target signal to perform the cross-PSD between it and the witness signals. Since the target signal is usually immersed in noise, and we cannot 
disentangle the two, we must use templates for the \uldmh target signal.
We injected templates of vector \uldmh signals (dark photons) in the data with different frequencies. Indeed, when we search for a signal in the data, we do not know which frequency or amplitude it could have. We only know it is almost monochromatic, and we can model it for different frequencies. We can therefore search for the right signal by calculating the residual (equation \ref{eq:WF-residual}) for many templates with different frequencies. The template should be in principle multiplied by $\epsilon_t$, the coupling value of the target template, to simulate different amplitudes, but we can see from equation \ref{eq:WF-residual} that the value of $\epsilon_t$ does not affect the residual, since it is contained in $X(\omega)$ and thus cancels out. For our purposes, then, we do not need to consider $\epsilon_t$ as a parameter for \uldmh searches with the Wiener filter. Instead, the signal strength, $\epsilon$,  will affect the residual, since it is contained in the measured $Y(\omega)$ signal and will change the SNR (see \ref{subsec:combin}). In Figure \ref{fig:Residual_vs_PSD}, we show that the residual width and shape coincide with the PSD signal width. By injecting signals with different $\epsilon$, we can see that the minimum residual value decreases with $\epsilon$; thus the SNR will increase. 
\begin{figure}
    \centering
    \includegraphics[width=0.49\textwidth]{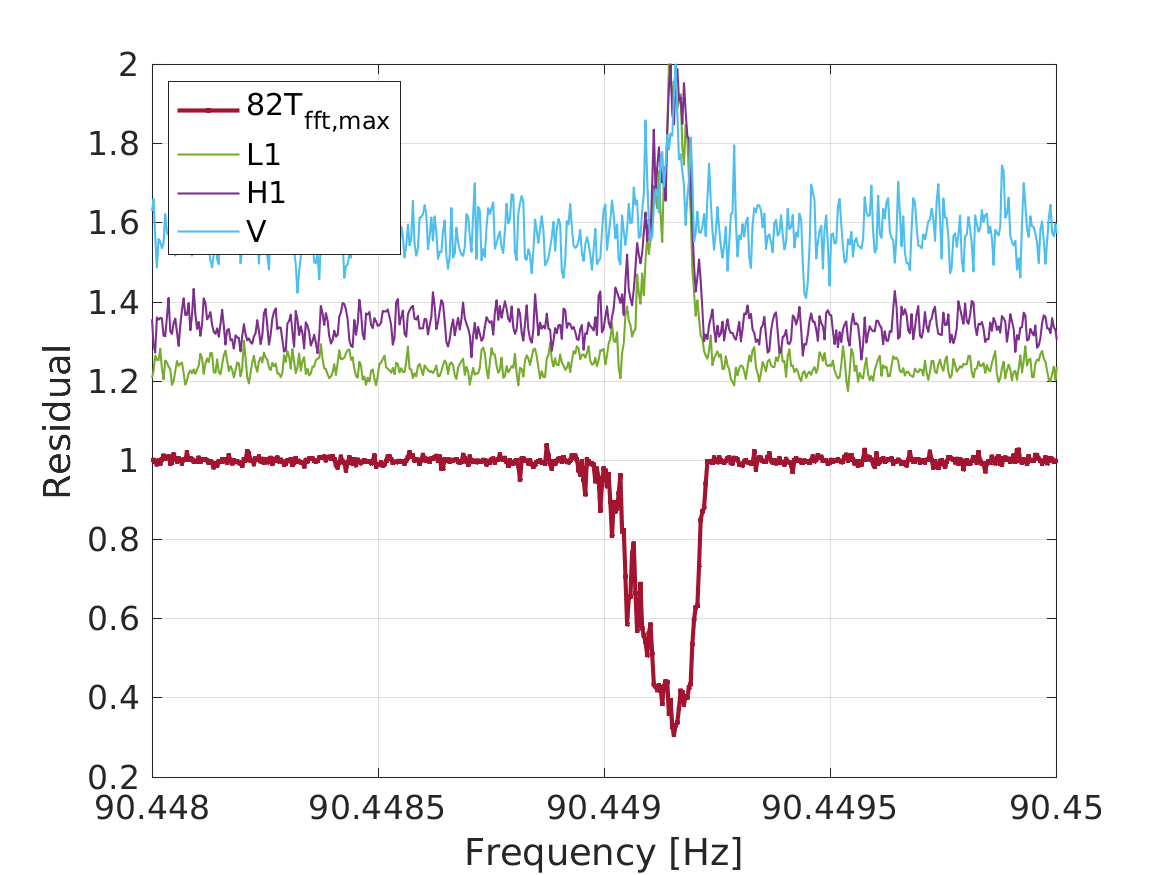}
    \caption{Comparison between the shape of the residual and the three detectors' PSDs for an injected dark photon \dmh signal. Each detector's PSD has been normalized and vertically shifted by 1 to easily compare it with the residual shape. In this plot, $\epsilon=10^{-22}$.}
    \label{fig:Residual_vs_PSD}
\end{figure}

\subsection{Detection statistic}
We would like to employ the Wiener filter as a validation method for the search of \uldmh signals. In this respect, our target signal will be the modeled \uldmh signal, while the witness sensors will be the \gwh detectors. Equation \ref{eq:WF-residual} provides an estimation for whether the target signal is present in the data: the lower the residual, the more likely it is that the target signal and data match. To distinguish potential \uldmh candidates from detector noise, we should employ the residual as a detection statistic. While we have so far understood that low residuals imply the presence of the target signal, we have not yet discussed how to quantify a ``low enough'' residual to claim that signal as ``significant''. In order to do this, we must construct a background distribution for the residual; i.e. what residuals would be we obtain if the target signal and that in the data are not a match? Once we have this distribution, we can place a threshold, $R_{\rm thr}$, below which we would consider a signal as interesting.

In practice, we injected 39 \uldmh signals into the LIGO Hanford, LIGO Livingston and Virgo detectors' data streams from the third observing run (O3). For each one of these data streams, we evaluated 99 residuals using a target signal with a {different frequency}. Hence, we have 3861 templates with which to construct our background distribution for the Wiener filter residual. Figure \ref{fig:Residual_histo} shows the histogram of all the residuals obtained, in two cases: (1) the signal we injected is strong ($\epsilon=10^{-18}$), and (2) the signal is very weak ($\epsilon=10^{-23}$). We can see that in both cases, residuals do not go below 0.70, meaning that in a real search, $R_{\rm thr}<0.7$, depending on what level of significance we wish to use to claim a candidate is interesting.  If a template returns a value for the residual above $R_{\rm thr}$, we can veto the outlier at that particular frequency; if it returns a residual value lower than $R_{\rm thr}$, it warrants further investigation. When we apply the Wiener filter to outliers in O3 \cite{LIGOScientific:2021odm}, as described in section \ref{subsec:fuO3}, we take $R_{\rm thr}=0.7$.

\begin{figure}[h!]
    \centering
    \includegraphics[width=0.49\textwidth]{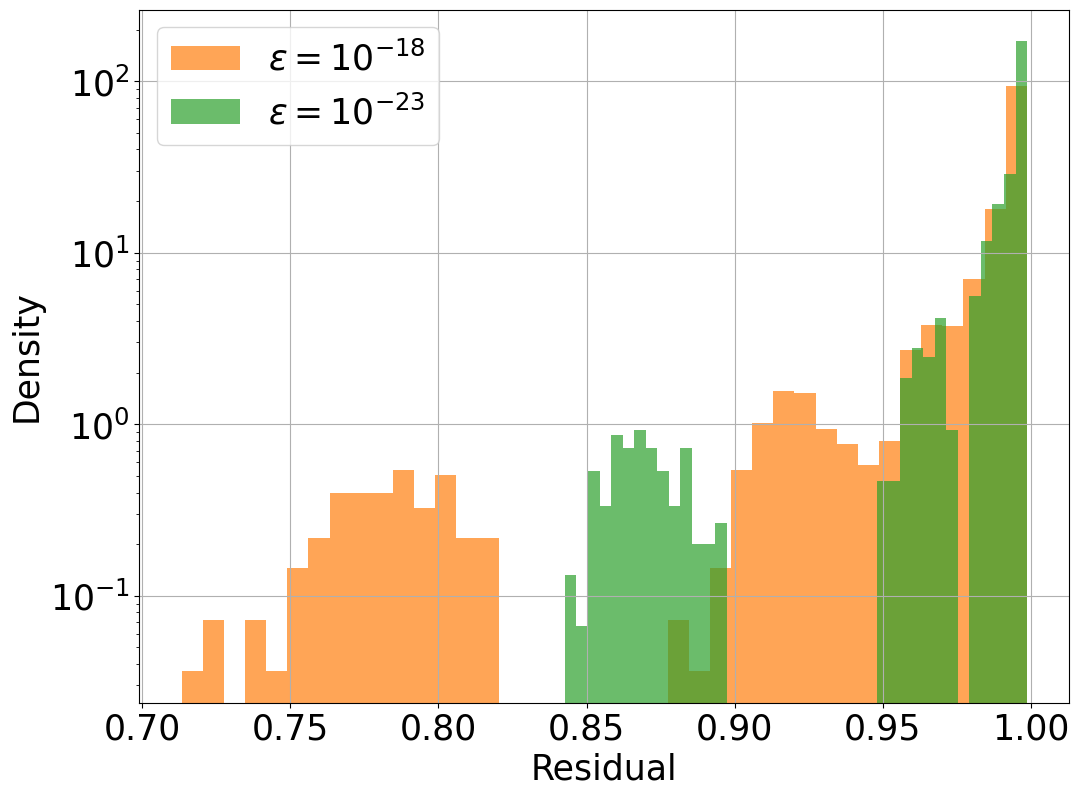}
    \caption{This plot provides a background distribution for the residual, to which we can compare the residuals from interesting \dmh candidates obtained in a real search. These residuals are calculated using 3861 dark-photon templates with frequencies between 40-1990 Hz that differed from the injected one ($f_0=90.449$ Hz, $m_A=3.741\times10^{-13}$ eV/$c^2$).  The mean and standard deviation of this residual distribution are 0.98 and 0.037, and 0.99 and 0.021 for the injected signal with $\epsilon=10^{-18}$ and $\epsilon=10^{-23}$, respectively.  }
    \label{fig:Residual_histo}
\end{figure}


\subsection{Application of Wiener Filter}

To apply the Wiener filter in the context of \uldmh searches in data from \gwh \ifos, we should understand what residuals to expect as a function of the strength of a potential \dmh signal. In a real search, we would then be able to map our residuals to strain amplitude of a signal, and, in the case of dark photons, the coupling $\epsilon$. {We inject the same signal with various coupling strengths into O3 Hanford, Livingston and Virgo data, and apply the Wiener filter to obtain residuals. We plot the residual value at the frequency of the \dmh signal in figure \ref{fig:Res_eps} as a function of $\epsilon$}. This plot gives an indication of the sensitivity of the Wiener filter method to extract weak \dmh signals from the data, though a full sensitivity study is beyond the scope of this work.

\begin{figure}
    \centering
    \includegraphics[width=0.49\textwidth]{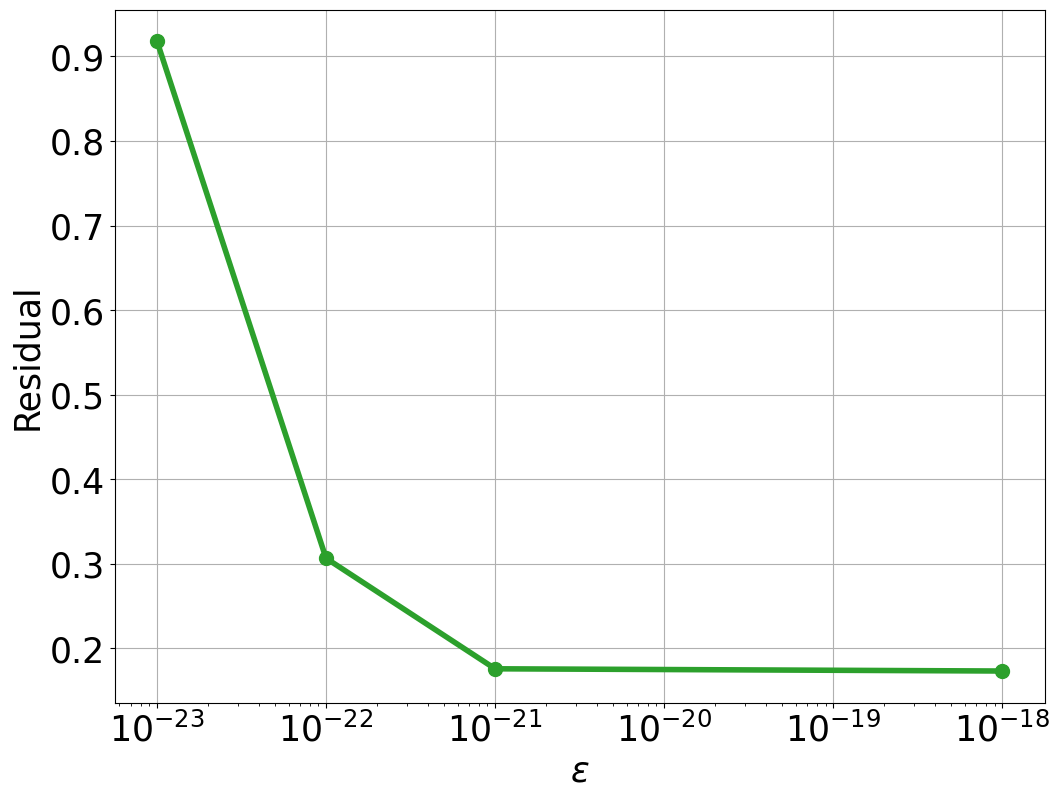}
    \caption{{Value of the minimum residual (corresponding to the frequency of the dark photon signal) versus the coupling strengths used to generate a dark photon signal with $\TFFT=82\tfftmax$. The injected signal frequency is: $f_0=90.449$ Hz ($m_A=3.741\times10^{-13}$ eV/$c^2$). }}
    \label{fig:Res_eps}
\end{figure}


We apply the frequency domain Wiener filter on data that has been Fourier transformed, and the length of each fast Fourier transform may impact the residual values we obtain. This is because as we increase $\TFFT$ beyond $\tfftmax\sim v_0^2/c^2 f \sim 10^{-6}f$ Hz, the maximum allowed fast Fourier Transform that would confine frequency modulations to one frequency bin, we start to observe more and more power spreading in the frequency domain near the signal frequency. Thus, in figure \ref{fig:Res_nfft}, we determine, for different dark photon coupling strengths, the values of the residual as a function of increasing $\TFFT$. Increasing $\TFFT$ appears to decrease the value of the minimum residual, thus helping to better distinguish the \uldmh signal from noise, which is especially true in the ``intermediate'' signal strength regime, i.e. $\epsilon=10^{-22}$. Eventually, however, the residual value tends to saturate around $\TFFT=40\tfftmax$, implying that we would not improve our estimation of the residual in a real search with $\TFFT>40\tfftmax$. 

\begin{figure}
    \centering
    \includegraphics[width=0.49\textwidth]{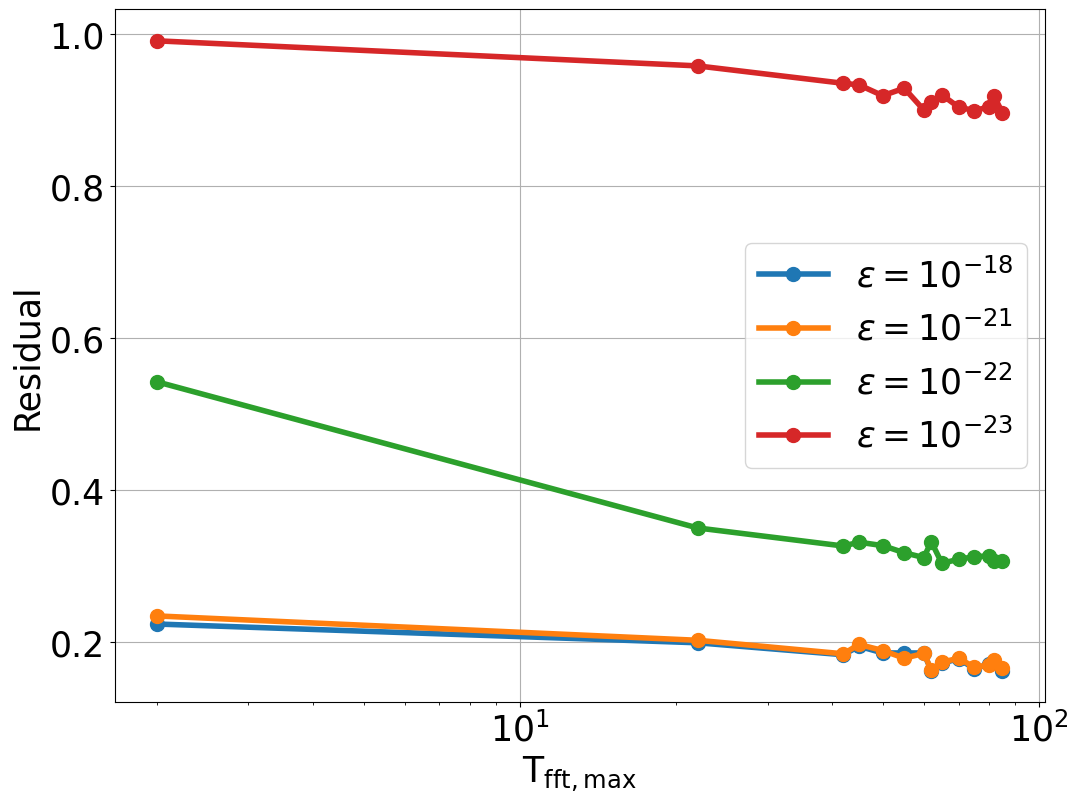}
    \caption{{Residuals versus the length of the fast Fourier Transform, in terms of the maximum allowed fast Fourier transform $\tfftmax$, for injected dark photon signals with different coupling strengths. We see that for very strong (orange, blue) and very weak signals (red), that the reduction in the residuals with increasing $\TFFT$ are marginal compared to an intermediate-strength signal (green). The injected signal frequency is: $f_0=90.449$ Hz ($m_A=3.741\times10^{-13}$ eV/$c^2$). } }
    \label{fig:Res_nfft}
\end{figure}

We can understand why this saturation of the residuals at a particular $\TFFT$ happens by looking at figure \ref{fig:Residual_vs_nfft}. Here, we show how the residual shape changes with different fast Fourier transform lengths for the green curve ($\epsilon=10^{-22}$) in figure \ref{fig:Res_nfft}. We observe very little difference in the residuals between curves with $\TFFT\geq 22\tfftmax$ because the power spreading that occurs beyond $22\tfftmax$ does not result in significant SNR loss.

\begin{figure}
    \centering
    \includegraphics[width=0.49\textwidth]{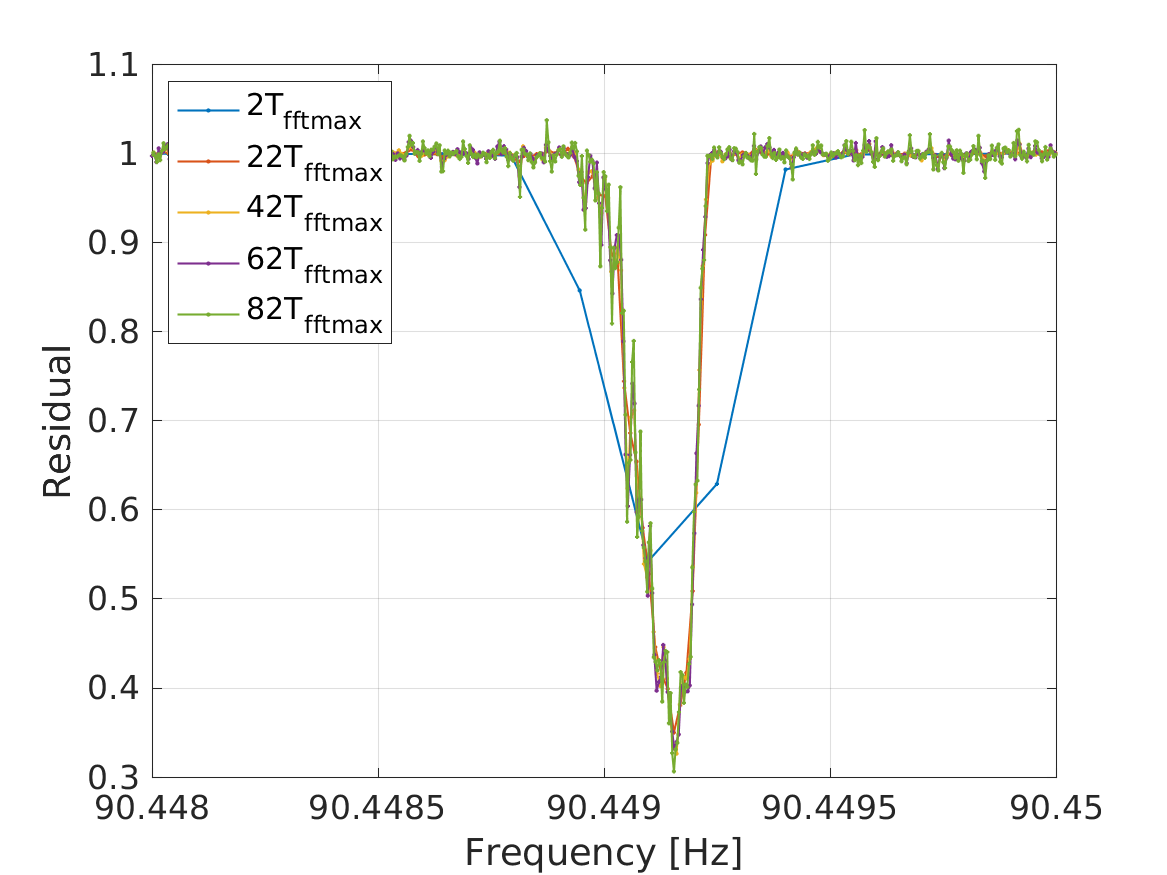}
    \caption{{Comparison between different fast Fourier transform lengths applied to an injected dark photon \dmh signal. When increasing $\TFFT$, the residual width around the dark photon signals tends to stabilize around that of the PSD (see figure \ref{fig:Residual_vs_PSD}). 
    In this plot, $\epsilon=10^{-22}$.}}
    \label{fig:Residual_vs_nfft}
\end{figure}

\subsection{Follow-up of candidates to rule out noise disturbances}
\label{subsec:fuO3}

The search for dark photon dark matter in LIGO/Virgo O3 data \cite{LIGOScientific:2021odm} using an excess power method \cite{Miller:2020vsl} returned 11 coincident outliers (i.e., particular frequencies) with critical ratios (our detection statistic) greater than five, implying that there could have been a signal at these frequencies. At the time, these outliers were vetoed by manually studying the spectra created with different fast Fourier Transform lengths, and showing that the outliers were in fact coincident with various noise disturbances in the data that became apparent at different fast Fourier Transform lengths. Instead of studying by eye each of the spectra, we can use the Wiener filter to veto these outliers, by showing that their residuals are approximately equal to one, which is what is expected when the target waveform (i.e., the dark photon or scalar \dmh signal) does not match the waveform returned by the analysis method.

Figure \ref{fig:o3outliers} shows the values of the residuals for each of the 11 outliers in the O3 analysis returned by the excess power method (table II in \cite{LIGOScientific:2021odm}), for the cases of applying the Wiener filter, with models of both a scalar and dark photon \dmh signal, to the data of each detector separately (diamonds) and then jointly (squares) for Hanford-Livingston (HL), Livingston-Virgo (LV) and Hanford-Virgo (HV). The color represents the values of the critical ratio returned by the original O3 search. Despite having high critical ratios, the values of the residuals for each of these outliers are nearly one, underscoring the effectiveness of the Wiener filter in vetoing such outliers even when the detection statistic takes a high value, i.e. greater than five, our threshold in the analysis.

\begin{figure}[htb]
    \centering
    \includegraphics[width=0.5\textwidth]{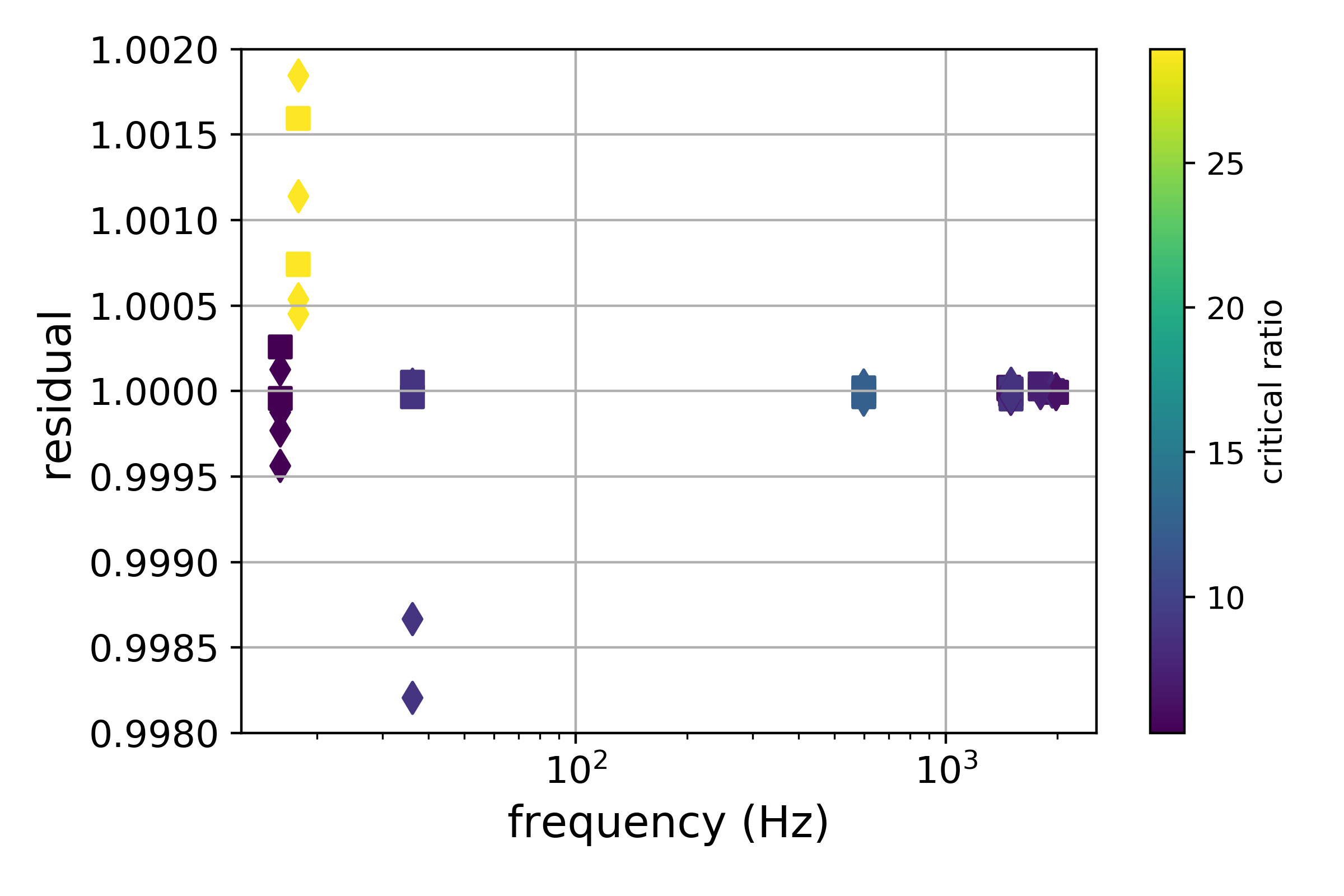}
    \caption{Residuals for each of the 11 outliers as a function of frequency, with the critical ratio (detection statistic in the excess power search in \cite{LIGOScientific:2021odm}) colored, obtained using both scalar and dark photon \dmh models. Squares indicate residuals calculated using a pair of detectors (HL, LV or HV), while diamonds denote residuals calculated for a single detector. Even for high values of the critical ratio, the residual values are very close to one, showing that the Wiener filter can successfully veto outliers that appear in the initial analysis \cite{Miller:2020vsl,LIGOScientific:2021odm}. We use $\TFFT=2\tfftmax$ here at each frequency.}
    \label{fig:o3outliers}
\end{figure}

\subsection{Ability to distinguish between dark matter interactions}

Because the Wiener filter requires a specific model, we can use it as a way to identify exactly what kind of \dmh signal is present in the data. In contrast to its deterministic counterpart, the matched filter, it should return a small value of the residual when the signal model matches what is present in the data, and a large value when the wrong model is used. To test this claim, we simulate both scalar and vector \dmh interaction signals, and inject them into real LIGO O2 Livingston data within the Band Sampled Data framework \cite{piccinibsd}. We then filter the data with a scalar injection, simulated based on codes available from \cite{Vermeulen:2021epa}, with a model for a vector signal that arises from an \uldmh particle with the same mass, and vice-versa, to determine whether the Wiener Filter will fail. For comparison, we also filter the data containing scalar and vector injections with themselves, to obtain the optimal, low value of the residual.

We show the frequency spectrum for injected scalar and vector \dmh signals when applying both scalar and vector signal filters in figure \ref{fig:oneinj}. Here, we see that the residuals obtained when filtering with the wrong model at the right frequency are $\sim 1$ while filtering with the correct signal gives values less than one. We note here that the injected vector (dark photon) \dmh signals are not that strong ($\epsilon=10^{-22}$), which explains the residual values around $\sim 0.7$. If we had injected the same or a stronger signal in O3 data, we would have obtained lower residuals, similar to those shown in figure \ref{fig:Res_eps}.

\begin{figure}[htb]
    \centering
    \includegraphics[width=0.5\textwidth]{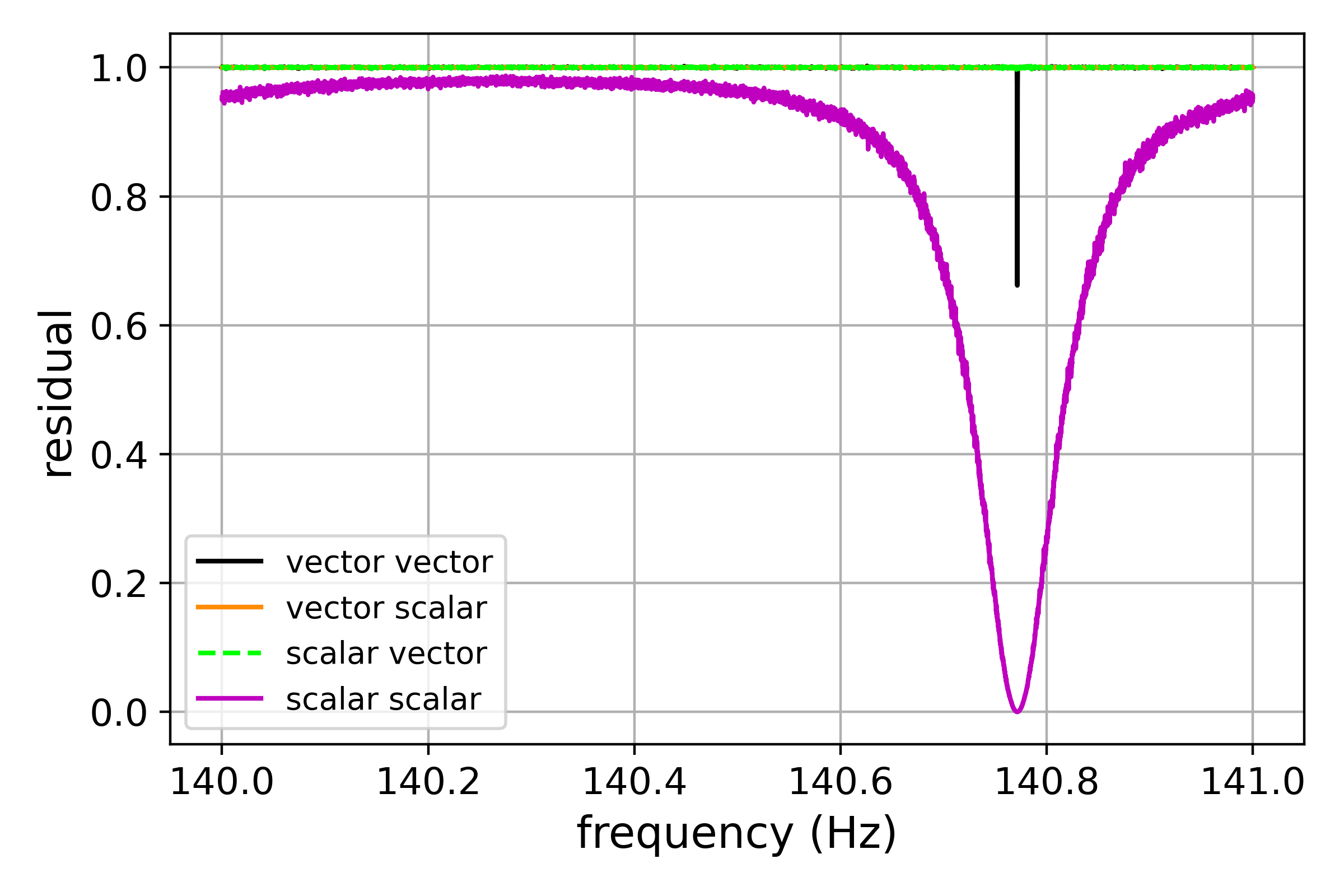}
    \caption{Residuals obtained for an injection with $f_0=140.77$ Hz ($m_\phi=m_A=5.82\times 10^{-13}$ eV/$c^2$) and a strain amplitude of $10^{-22}$ and $10^{-24}$ for the scalar and vector injected signals, respectively. The residuals of filtering a scalar injection with a vector model, and vice-versa, are so close to one and are indistinguishable. Here, $\TFFT=2\tfftmax$.}
    \label{fig:oneinj}
\end{figure}

We now expand our analysis to a sample of randomly chosen injections. In figure \ref{fig:inj_resids}, we plot the residual as a function of injected signal frequency for 10 injections. Here, we see that scalar injections filtered by vector models, and vice-versa, give values of the residual close to 1, while filtering the data with the exact model of the signal injected gives residual values that are more indicative of a signal (based on figure \ref{fig:Residual_histo}).

\begin{figure*}[ht!]
     \begin{center}
        \subfigure[ ]{%
            \label{fig:inj_resids}
            \includegraphics[width=0.5\textwidth]{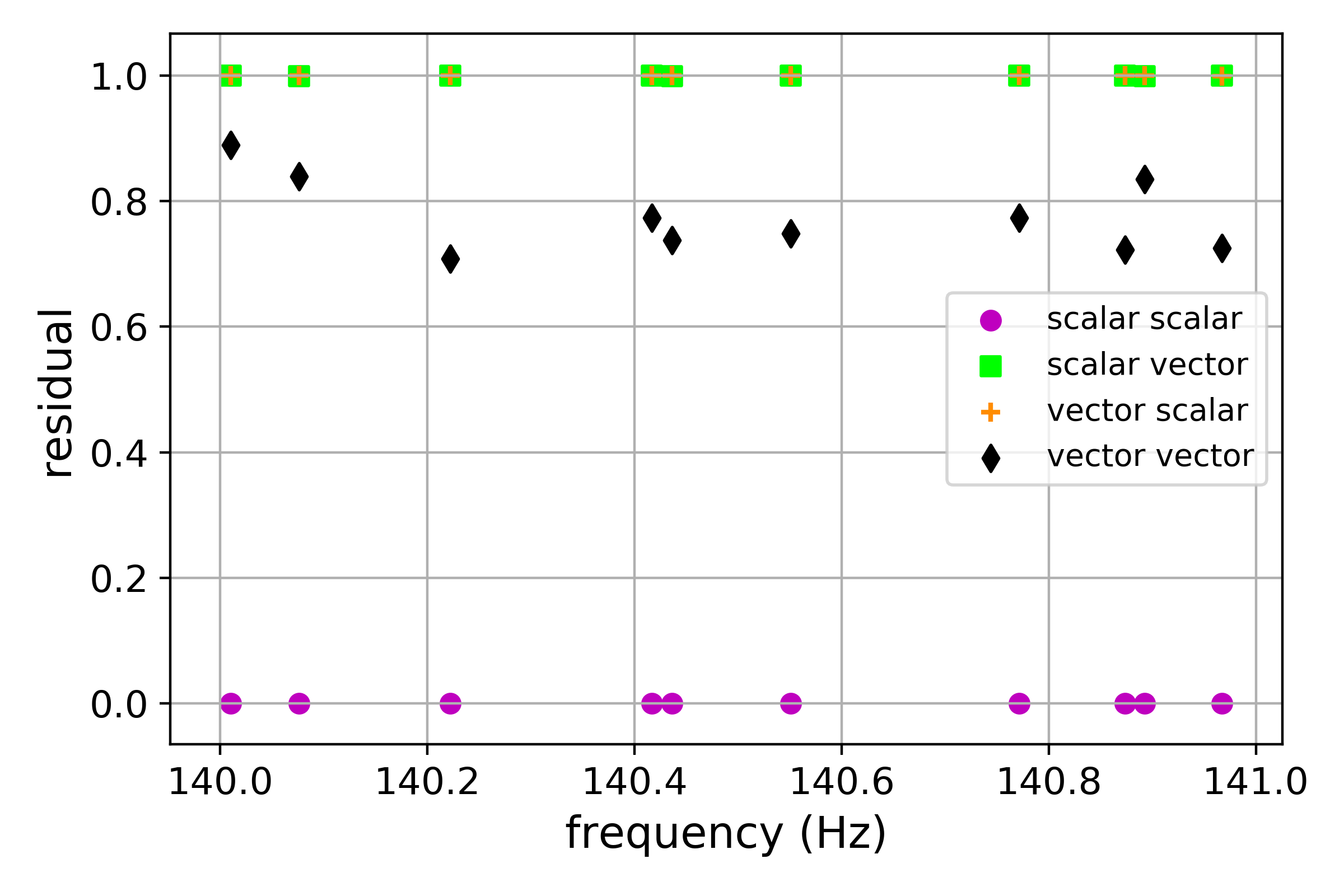}
        }%
        \subfigure[]{%
           \label{fig:confmat}
           \includegraphics[width=0.5\textwidth]{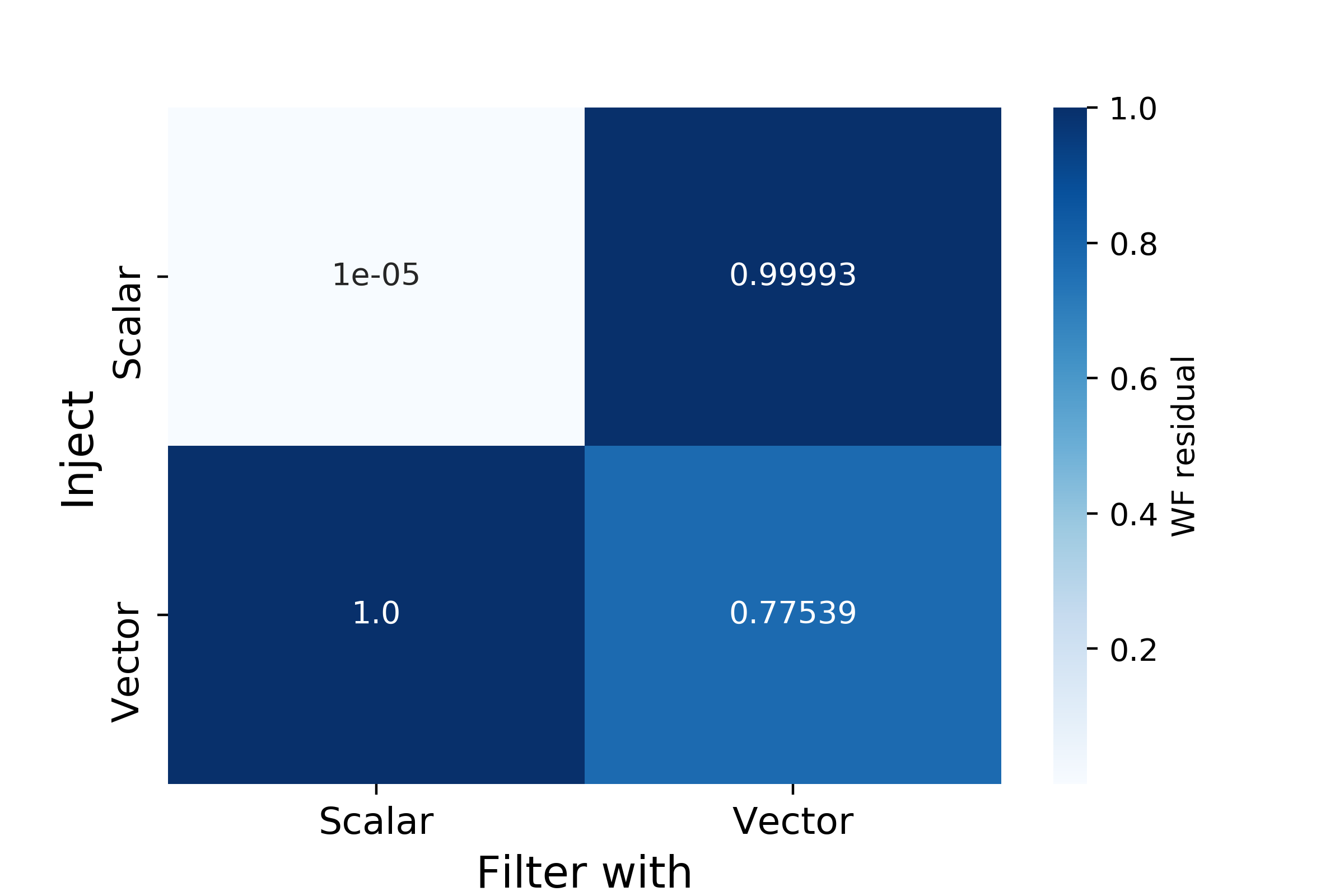}
        }\\ 
    \end{center}
    \caption[]{%
   Left: Residuals obtained from filtering 10 injected scalar or vector \dmh signals with scalar or vector waveforms. The strain amplitude of the scalar and vector signals are $\sim 10^{-22}$ and $\sim 10^{-25}$ ($\epsilon=10^{-22})$, respectively, which explains why the residuals for the ``scalar filtered with scalar'' case are so much lower than the ``vector filtered with vector'' case. Right: Average of residuals obtained from filtering 10 scalar and vector injections with scalar and vector models. Lower values of the residuals imply that, on average, the injected signal is well-recovered, while residuals close to unity indicate that the injected signal is not found. We have employed $\TFFT=2\tfftmax$ here.
     }%
\end{figure*}

We also plot, in figure \ref{fig:confmat}, a ``confusion matrix'' for the four permutations of injecting and filtering with scalar and vector signals. The numbers shown here are the average residuals across 10 scalar or vector injections when filtering with either the scalar or vector waveforms.  Figure \ref{fig:confmat} illustrates conceptually the ability of the Wiener Filter to distinguish between different \dmh models.




\section{Conclusions}\label{sec:concl}

{In this paper, we have shown that the Wiener filter can recover scalar and vector \dm that interact directly with \gwh \ifos. Within the context of method development, we have derived a detection statistic that can be used in a real search to quantify whether a \dmh signal is present in the data. We have shown what residuals to expect for various dark photon \dm coupling strengths, and have computed how the residuals would improve with increasing fast Fourier transform length. 
We have also determined that the Wiener filter can be employed as a robust follow-up method in \uldmh searches, which confirm or deny the presence of outliers with high detection statistic values. 
Finally, we have shown that the Wiener filter can distinguish between different \dmh signal models.}

{Our work represents the first step towards the inclusion of a robust follow-up method to determine the existence of \uldm, and provides a proof-of-concept study on the efficiency of the Wiener filter in \uldmh searches. Future work will include performing a comprehensive sensitivity study for various \uldmh signals with different amplitudes and using different fast Fourier transform lengths, the development of a generic ``template'' based solely on the Maxwell-Boltzmann velocity distribution that could enhance the signal in a model-independent way before the application of the Wiener filter, an estimate of the computational cost to employ the Wiener filter as a complete method for \uldmh searches, and a comparison with standard techniques already employed to detect \uldm. We will also expand the Wiener filter to search for tensor \dm \cite{Marzola:2017lbt}, which could arise from modifications to gravity, e.g. bimetric gravity \cite{Hassan:2011zd,Schmidt-May:2015vnx}, due to an additional spin-2 particle that would act as \dm. It has already been shown that current and future detectors, such as Cosmic Explorer \cite{Reitze:2019iox} and Einstein Telescope \cite{punturo2010einstein} on the ground, and DECIGO \cite{Kawamura:2020pcg}, LISA \cite{Babak:2017tow}, and TianQin \cite{Luo:2015ght} in space, could also be sensitive to tensor \dm \cite{Armaleo:2020efr}, and in general, greatly lower the noise floor, enhance our sensitivity these types of \dmh interaction signals, and cover an even lighter range of \dmh masses, i.e. $\mathcal{O}(10^{-17}-10^{-15})$ eV/$c^2$}. The future is bright for Wiener filter-based methods to not only follow-up candidates of other searches but also directly search for \uldmh interactions on earth and in space.


\section*{Acknowledgments}

This material is based upon work supported by NSF's
LIGO Laboratory which is a major facility fully funded
by the National Science Foundation.

We thank Ornella J. Piccinni and Sergio Frasca for the development of the Band Sampled Data framework that allowed us to easily perform simulations of \dmh signals. 

We thank Bernard Whiting for additional discussions regarding matched filtering and the physics of the dark photon dark matter signal, and the Amaldi Research Centre at Sapienza Università di Roma for support.

This research has made use of data, software and/or web tools obtained from the Gravitational Wave Open Science Center (https://www.gw-openscience.org/ ), a service of LIGO Laboratory, the LIGO Scientific Collaboration and the Virgo Collaboration. LIGO Laboratory and Advanced LIGO are funded by the United States National Science Foundation (NSF) as well as the Science and Technology Facilities Council (STFC) of the United Kingdom, the Max-Planck-Society (MPS), and the State of Niedersachsen/Germany for support of the construction of Advanced LIGO and construction and operation of the GEO600 detector. Additional support for Advanced LIGO was provided by the Australian Research Council. Virgo is funded, through the European Gravitational Observatory (EGO), by the French Centre National de Recherche Scientifique (CNRS), the Italian Istituto Nazionale della Fisica Nucleare (INFN) and the Dutch Nikhef, with contributions by institutions from Belgium, Germany, Greece, Hungary, Ireland, Japan, Monaco, Poland, Portugal, Spain.

Computational resources have been provided by the supercomputing facilities of the Université catholique de Louvain (CISM/UCL) and the Consortium des Équipements de Calcul Intensif en Fédération Wallonie Bruxelles (CÉCI) funded by the Fond de la Recherche Scientifique de Belgique (F.R.S.-FNRS) under convention 2.5020.11 and by the Walloon Region.

We also wish to acknowledge the support of the INFN-CNAF computing center for its help with the storage and transfer of the data used in this paper.

The authors gratefully acknowledge the support of the NSF, STFC, INFN and CNRS for provision of computational resources.

All plots were made with the Python tools Matplotlib \cite{Hunter:2007ouj}, Numpy \cite{Harris:2020xlr}, and Pandas \cite{mckinney-proc-scipy-2010,reback2020pandas}.

A.L.M. is a beneficiary of a FSR Incoming Post-doctoral Fellowship. 

We would like to thank all of the essential workers who put their health at risk during the COVID-19 pandemic, without whom we would not have been able to complete this work.

\bibliographystyle{apsrev4-1}
\bibliography{biblio}

\end{document}